\documentclass[aps,prb,twocolumn]{revtex4-2}

\usepackage{float}
\usepackage{amsmath}
\usepackage{graphicx}
\usepackage{caption}
\usepackage[export]{adjustbox}
\usepackage{hyperref}
\usepackage{color}
\usepackage{mathrsfs}
\usepackage{comment} 
\usepackage{braket}
\usepackage[normalem]{ulem}
\usepackage{physics}

\begin{document}

\title{Thermodynamics of the disordered Hubbard model studied via numerical linked-cluster expansions}

\author{Jacob Park}
\author{Ehsan Khatami}
\email[Corresponding author: ]{ehsan.khatami@sjsu.edu}
\affiliation{Department of Physics and Astronomy, San Jos\'{e} State University, San Jos\'{e}, California 95192, USA}

\date{\today}

\begin{abstract}
The interplay of disorder and strong correlations in quantum many-body systems remains an open question. 
That is despite much progress made in recent years with ultracold atoms in optical lattices to better 
understand phenomena such as many-body localization or the effect of disorder on Mott metal-insulator transitions. 
Here, we utilize the numerical linked-cluster expansion technique, extended to treat disordered quantum 
lattice models, and study exact thermodynamic properties of the disordered Fermi-Hubbard model on the 
square and cubic geometries. We consider box distributions for the disorder in the onsite 
energy, the interaction strength, as well as the hopping amplitude and explore how energy, double occupancy,
entropy, heat capacity and magnetic correlations of the system in the thermodynamic limit evolve as the strength of 
disorder changes. We compare our findings with those obtained from determinant quantum Monte Carlo 
simulations and discuss the relevance of our results to experiments with 
cold fermionic atoms in optical lattices.
\end{abstract}

\maketitle

\section{Introduction}

The interplay between electronic correlations and quenched (static) disorder is not well understood. 
From the experimental point of view, condensed matter experiments aiming at realizing either the 
Mott transition or Anderson localization in real materials have to worry about the presence of 
disorder in case of the former, or correlation effects in case of the latter since Coulomb 
interaction will always be present. For example, the expected first-order Mott transition in 
phosphorus-doped silicon upon increasing dopants is found to be continuous due to the random 
distribution of dopants in the so-called Mott-Anderson transition~\cite{g_thomas_81}. 
Experiments of disordered two-dimensional (2D) electron gases done in silicon metal-oxide-semiconductor 
field-effect transistors~\cite{s_kravchenko_94} find weak localization, but also find 
several other phenomena, such as a region of linear dependence of resistivity on temperature 
and a sharp drop in resistivity at low temperatures that may be explained only if one takes 
into account other factors such as the range of scattering centers, and electronic correlations 
in cleaner samples. Later studies~\cite{a_punnoose_05,a_anissimova_07} shed 
more light on the role of interaction in the metal-insulator transition and quantum criticality in these materials.

The presence of quenched disorder in strongly correlated materials exhibiting 
unconventional properties also confirms the need to treat 
disorder and electronic correlations on the same footing in order to 
be able to describe the complex and competing phases~\cite{gulacsi,w_fu_16}. 
Recent examples are the interplay 
of charge density wave order and high-temperature superconductivity in
cuprates~\cite{m_leroux_19} or the spin liquid behavior in 
herbertsmithides~\cite{s_lee_07,f_bert_07}.

Encouraged by an unprecedented control and measurement possibilities in experiments with 
ultra-cold atomic gasses~\cite{i_bloch_08}, pure Anderson localization of matter waves 
was first realized in 2008,~\cite{j_billy_08,g_roati_08} using
disorder potentials introduced  through laser speckles~\cite{j_lye_05}, or a quasi-periodic  
lattice potential~\cite{s_aubry_80} in Bose-Einstein condensates trapped in one dimension.
These experiments generated a lot of interest in the community and  later inspired the 
experimental realization of the disordered Bose-Hubbard 
model~\cite{b_gadway_11,m_white_09,m_pasienski_10}. 

Presently, an exciting new frontier is the exploration of many-body localization
through the simulation of disordered models in optical lattices~\cite{m_schreiber_15}, and the idea that disorder 
may in fact help achieve lower temperatures in experiments~\cite{t_paiva_15}. In a 
pioneering work, Kondov et al.~\cite{s_kondov_15} studied the Fermi-Hubbard model experimentally 
in the limit of strong correlations and found that by increasing the interaction strength 
the system undergoes an insulator to metal transition, and that for larger disorder 
strengths the onset of the transition is moved to larger interactions. Consistent 
with many-body localization prediction, there is also a lack of thermally activated conductivity. 
More recently, out-of-equilibrium properties of the disordered Fermi-Hubbard model 
in three dimensions (3D) has also been studied across a range of disorder and interaction 
strengths, where phenomena such as ``bad metal" and a disorder-induced pseudogap 
were observed~\cite{w_morong_16}. 

While phenomenological theory and approximations 
can describe some of these observations, the need for reliable exact results of the microscopic 
model is greater than ever. In light of the great progress made towards our understanding 
of spin and charge correlations in the clean 2D and 3D Fermi-Hubbard models emulated
in optical lattice experiments through cross comparison of results with exact 
numerical solutions in recent 
years~\cite{r_hart_15,m_parsons_16,l_cheuk_16,p_brown_17,a_mazurenko_17,c_chiu_18,p_brown_18,m_nichols_19,m_gall_21}, 
the new progress in preparing and manipulating 
disordered quantum lattice models calls for highly-precise and readily available 
numerical solutions of the corresponding models in temperature ranges relevant
to experiments.

Among the methods that deal with the disordered Fermi-Hubbard model are the determinant quantum Monte Carlo 
(DQMC)~\cite{r_blankenbecler_81,m_ulmke_98,p_denteneer_99,d_heidarian_04,k_bouadim_11,e_lahoud_14,t_paiva_15},
the Monte Carlo mean-field approximation~\cite{a_mukherjee_14}, 
the density matrix renormalization group~\cite{s_white_92,j_wernsdorfer_11}, and dynamical mean-field theory
(DMFT)-based methods~\cite{a_georges_96,m_hettler_98,m_ulmke_98} extended to incorporate disorder, such as
the statistical DMFT~\cite{v_dobrosavljevic_97}, the DMFT+Sigma approach~\cite{m_sadovskii_05,e_kuchinskii_15}, which
dresses the local Green's function by an additional approximate self energy due to interactions 
outside of the DMFT (here, disorder), or the typical-medium dynamical cluster approximation~\cite{c_ekuma_14,h_terletska_14},
in which the cluster density of states is replaced by its typical value, but the local part of
the typical density of states is explicitly separated out and geometrically averaged over disorder 
configurations. Exact diagonalization with disorder averaging has also been used to study 
the equilibrium and non-equilibrium properties of the Hubbard models~\cite{r_kotlyar_01,r_mondaini_15}.

While these techniques can access low-temperature properties and even explore quantum phase
transitions of the disordered Hubbard models, each suffers from one or more limitation
that may prevent it from being the ideal candidate for the characterization of the systems studied in 
optical lattice experiments in some parameter region. The other major issue with methods that take the 
disorder average randomly is the introduction of statistical errors associated with disorder 
averaging, which can introduce significant fluctuations in the calculated properties even at 
intermediate temperatures. The QMC-based methods are also better suited for systems with only weak- 
to intermediate-strength interactions; as we will see later they can run into technical difficulties
in strong-coupling regions of the Hubbard models. Moreover, comparisons to experimental 
data often requires fast calculations of thermodynamic properties for a wide range of model
parameters, temperatures, and densities, something that takes considerable time and computational 
resources to achieve with methods that do not have access to the full energy spectrum.

In this paper, we use the numerical linked-cluster expansion (NLCE)~\cite{M_rigol_06,b_tang_13b}, 
extended to treat random disorder~\cite{m_mulanix_18}. The main advantage of the NLCE is the fact that it
yields exact finite-temperature results for the Hubbard model {\it in the thermodynamic limit} 
(no finite-size or statistical errors). Moreover,
similar to ED of finite clusters, one can obtain all the properties for a set of model parameters in a 
single run on an arbitrarily dense temperature or density grid. The process is fast and 
allows one to perform a systematic study of thermodynamic properties of the disordered model in 2D and 3D. 
While highlighting the effectiveness of these features here, it should be noted that 
the main weakness of the NLCE over most other numerical methods 
is its limitation in reaching low temperatures; the convergence of the series expansion 
is typically lost at a finite temperature that in general depends on the model and its parameters.

We separately consider site, interaction, and hopping disorders with a range of strengths
and monitor properties such as the average energy, double occupation fraction, heat capacity, 
entropy and spin correlations as a function of temperature. We find that these properties
are much more sensitive to disorder in the chemical potential than disorder in the onsite 
interaction strength. At half filling, the former strongly suppresses the magnetic correlations
by suppressing moment formation and 
promotes a state in which particles are localized at sites with significantly lower
chemical potential, as evidenced by the enhancement of the fraction of doubly-occupied sites 
as the temperature is lowered. The bond disorder hinders the NLCE's ability
to access low temperatures, however, we find evidence for the formation of dimers on 
strong bonds at low temperatures. We also present results from the DQMC after disorder realization
averaging, which show good agreement with our exact NLCE results and expose the 
strengths and weaknesses of each method in different parameter regions.

\section{Models}

We study the Fermi-Hubbard model~\cite{n_mott_49,j_hubbard_63} with 
random box disorder introduced to its various parameters through the 
following Hamiltonian:
\begin{eqnarray}
H=&-&\sum_{\left<ij\right> \sigma}t_{ij}c^{\dagger}_{i\sigma}
c^{\phantom{\dagger}}_{j\sigma}\nonumber\\
&+& \sum_i U_i \left( n_{i\uparrow}-\frac{1}{2}\right )\left (n_{i\downarrow}-\frac{1}{2}\right )\nonumber\\
&-&\sum_i \mu_i n_i,
\label{eq:H}
\end{eqnarray}
where $c^{\phantom{\dagger}}_{i\sigma}$ ($c^{\dagger}_{i\sigma}$)
annihilates (creates) a fermion with spin $\sigma$ on site $i$,
$n_{i\sigma}=c^{\dagger}_{i\sigma} c^{\phantom{\dagger}}_{i\sigma}$ is
the number operator, and $\left<.. \right>$ denotes nearest neighbors. 
We consider random onsite Coulomb interactions $U_0-\Delta_U<U_i<U_0+\Delta_U$, 
hopping integrals, $t_0-\Delta_t <t_{ij}<t_0+\Delta_t $, or onsite energies 
$\mu_0 - \Delta_\mu<\mu_i< \mu_0 + \Delta_\mu$ drawn from a uniform distribution.
$\Delta_U, \Delta_t, $ and $\Delta_\mu$ are strength of disorder for the Coulomb 
interaction, hopping amplitude, and the onsite energy, respectively. We set
$t_0=1$ as the unit of energy, and except when studying the effect of $U_0$,
keep $U_0=8$ fixed throughout the paper. In any given calculation,
we choose only one of the three disorder strengths to be nonzero. We work 
mostly with the half filled model in 2D on a simple square lattice but
also study the model away from half filling and on the 3D cubic lattice with 
interaction and onsite energy disorders.

\section{Methods}

We use the NLCE for disordered quantum lattice models as described in detail in Ref.~\cite{m_mulanix_18}.
In the NLCE, one expresses an extensive property of the model as a series in terms of ``reduced properties" 
associated with all connected (linked) clusters, $c$, that can be embedded in the lattice $\mathscr{L}$,
\begin{equation}
\label{eq:nlce_exp}
P(\mathscr{L})= \sum_{c} W_P(c).
\end{equation}
The reduced properties, $W_P(c)$, are in turn computed using the inclusion-exclusion principle:
\begin{equation}
\label{eq:nlce_weight}
W_P(c) = P(c) - \sum\limits_{s \subset c} W_P(s),
\end{equation}
where $s$ is a cluster that can be embedded in $c$ (a sub-cluster of $c$) and $P(c)$ is the 
property for the finite cluster $c$ calculated at finite temperature exactly 
using full numerical diagonalization.

When $\mathscr{L}$ represents an infinite lattice, it is more straightforward to work with 
the normalized quantity $\lim_{\mathscr{L}\to \infty}P(\mathscr{L})/\mathscr{L}$ (we have 
taken $\mathscr{L}$ to represent the number of sites in the lattice too). In that case,
reduced properties only for those clusters not related via translational symmetry need to be included.
Moreover, point group symmetries of the underlying lattice can be used to further optimize
the calculations by considering a multiplicity factor and contributions from clusters that
are topologically distinct and are not related through point group symmetries. 

In the case
of the infinite lattice, one is forced to truncate the series and include contributions from finite 
clusters only up to a certain size due to limitations either on the number of clusters 
that have to be solved or time and memory requirements for diagonalizing the largest 
clusters in the expansion. Therefore, in finite-temperature calculations, one typically 
loses convergence below a temperature where correlations in the system grow beyond 
the order of the largest clusters considered. This temperature depends on the model, its parameters
and the order of the expansion. Details of the NLCE algorithm can be found in Ref.~\onlinecite{b_tang_13b}.

Here, we use the site expansion in which order $l$ means we have included contributions
from all clusters with up to $l$ sites. Although numerical resummation algorithms can
be used to extend the region of convergence of the NLCE, due to the relatively small 
number of terms kept in this study for the Hubbard model in the presence of disorder, 
we have not explored this possibility and have restricted ourselves to working with the 
raw expansion.

As described in Refs.~\onlinecite{r_singh_86,b_tang_15,b_tang_15b,m_mulanix_18}, the application of 
NLCEs can be extended to disordered systems by replacing $P(c)$ for finite clusters in the
above equations by their disorder realization averaged values. However, in Ref.~\onlinecite{m_mulanix_18},
the authors discuss that the straightforward approach of averaging properties using randomly 
generated realizations may cause the NLCE to break down due to the propagation of statistical 
errors, unless the error bars can be driven down to the order of the machine precision~\cite{t_devakul_15}. 
Therefore, a more systematic and statistical-error-free procedure was introduced in which the limit of random disorder 
was approached by increasing the number of disorder modes, $m$, in a discrete ``multi-modal" 
distribution, allowing finite sums over disorder realizations to be taken exactly. It was found that
with an efficient choice of mode locations in the box distribution, the convergence in the number 
of disorder modes could be fast, typically achieved with $m\lesssim 6$ at the lowest temperatures 
the NLCE converges~\cite{m_mulanix_18}.

Here, we adopt the same algorithm and apply the method to the disordered Fermi-Hubbard models 
to obtain exact finite-temperature results for the thermodynamic quantities in the limit of an infinite lattice.
To compare our results with those obtained via DQMC simulations of finite clusters,
we employ the QUEST package~\cite{quest} and, unless stated otherwise, average 
expectation values obtained on a $10\times 10$ periodic cluster in each case over at least a 
hundred random disorder realizations. The imaginary time step in DQMC is chosen to be $0.01$ 
at $T\ge 2$ for all $U_0$ and $0.1$ ($0.05$) at $T<2$ for $U_0=4$ or $8$ ($U_0=16$). The comparison 
allows us to gauge any systematic errors as well as fluctuations due to the disorder that may exist in the latter. 
After discussing the convergence in the number of modes in the NLCE for the case of disorder in the onsite energies,
we present results for a range of disorder strengths for each of the three types of disorder in the model.

\section{Results}

\subsection{Convergence in disorder modes}

 \begin{figure}[t]
	\includegraphics[width=1\linewidth]{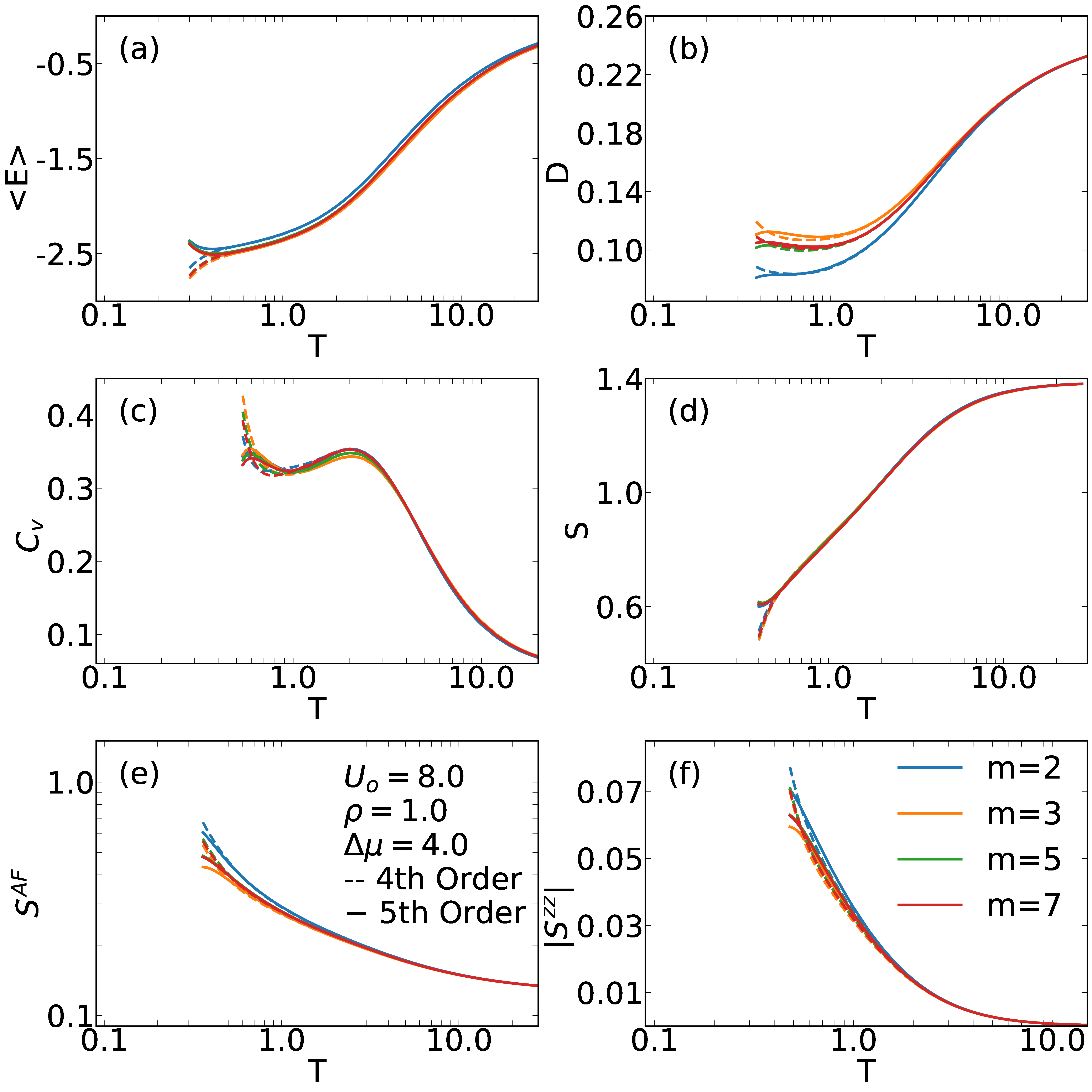}
 \caption{\label{fig:initial}   
 NLCE results for the (a) average energy, (b) double occupancy, 
 (c) specific heat, (d) entropy, (e) antiferromagnetic structure factor, and (f) absolute value of 
 nearest-neighbor spin correlation vs temperature for the Hubbard model at half filling with 
 onsite potential disorder and different number of disorder modes in the discrete 
 distribution. The convergence of the series is lost below some temperature for the 
 shown fourth and fifth orders, with fourth order results plotted as dashed lines and fifth order 
 plotted as solid lines. The interaction strength is $U_0=8$ and the strength of 
 the chemical potential disorder is fixed at $\Delta_\mu=4$.}
 \end{figure}
 
We first examine the dependence of thermodynamic quantities on the number of disorder 
modes in the method at various temperatures. We use the average energy $E=\expval{H}$, 
double occupancy fraction $D=\expval{n_{\uparrow}n_{\downarrow}}$, spin correlations, 
heat capacity $C_v$, and entropy $S$. The latter two are calculated without performing 
a numerical derivation or integration, rather, by using the knowledge of the partition 
function ($Z$) and other correlation functions, which are available in the NLCE within 
machine precision for a given cluster and disorder realization~\cite{e_khatami_12b}:
\begin{equation}
S=\ln(Z)+\frac{\expval{H} -\mu \expval{n}}{T},
\end{equation}
and
\begin{eqnarray}
C_v &=&\frac{1}{T^2}\left[\expval{\Delta H^2}-\frac{
\left (\expval{H n}-\expval{H}\expval{n}\right)^2}{
\expval{\Delta n^2}}\right],
\end{eqnarray}
where $\rho=\expval{n}=\expval{n_{\uparrow}+n_{\downarrow}}$ is the average density, 
and $\expval{\Delta H^2} = \expval{H^2}-\expval{H}^2$, and similarly 
$\expval{\Delta n^2} = \expval{n^2}-\expval{n}^2$. At half filling,
where $\rho=1$, the expression for $C_v$ reduces to the more familiar 
$\expval{\Delta H^2}/T^2$. 
For magnetic properties, we study the nearest-neighbor spin correlations along $z$,
$S^{zz}=\frac{1}{M}|\langle\sum_{\bf r}S^z_i S^z_{i+{\bf r}}\rangle|$, where 
$S^z=(n_{\uparrow}-n_{\downarrow})/2$ and the sum 
runs over the $M$ nearest neighbors of site $i$, and the antiferromagnetic structure 
factor $S^{AF}=\expval{\left( \sum_{i} \phi_i S^z_i \right)^2}$, where
the phase $\phi_i$ alternates between $\pm 1$ on neighboring sites. 

\begin{figure}[t]
\includegraphics[width=1\linewidth]{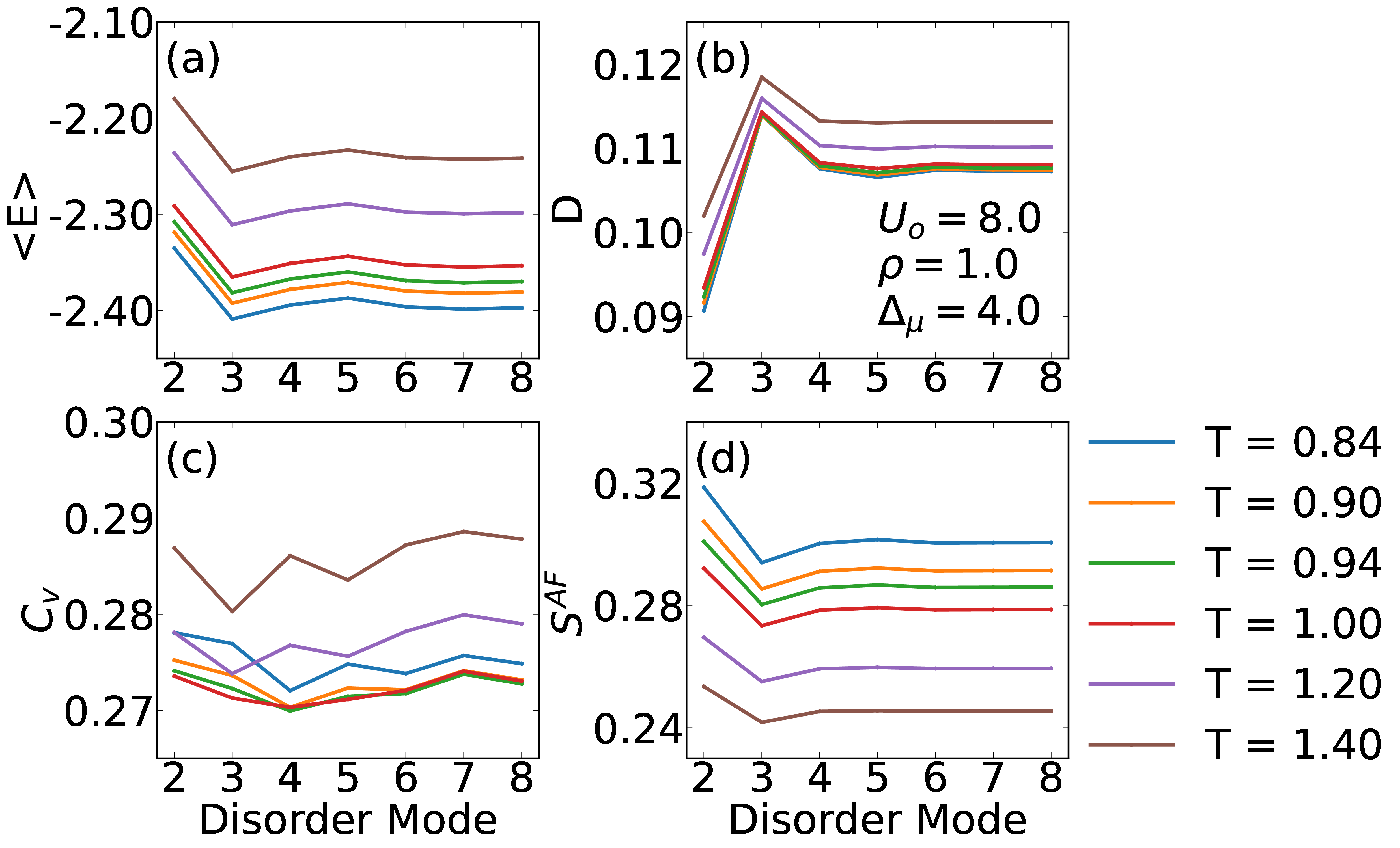}
\caption{\label{fig:vsm} Convergence in the number of disorder modes at various temperatures. 
Same (a) average energy, (b) double occupancy, (c) specific heat, and (d) antiferromagnetic 
structure factor as in Fig.~\ref{fig:initial} plotted here vs $m$ at a few select 
temperatures above and below $t_0$. The values are from NLCE results in the fifth order.}
\end{figure}

Consistent with previous results for magnetic models~\cite{m_mulanix_18}, we find 
that the convergence of these properties with increasing the number of modes is fast, and is achieved 
typically with four to six modes at temperatures available to the NLCE. Figure~\ref{fig:initial} shows the 
results for a system at half filling ($\mu_0=0$) and a disorder in the chemical potential 
with the strength $\Delta_\mu=4t_0$ from the fourth (dashed lines) and fifth (solid lines) orders of the NLCE. 
Up to seven disorder modes ($m=7$) are shown for each case. First, we observe that the NLCE for 
any individual $m$ is converged generally for temperatures
above $t_0$ with the energy, entropy and the antiferromagnetic structure factor showing an extended region of 
convergence down to $T\sim 0.5t_0$ for most values of $m$. Second, we see that the curves for $m=5$ 
and $m=7$ are almost indistinguishable in the temperature ranges shown for all of the quantities,
except for $C_v$ in Fig.~\ref{fig:initial}(c), where significant differences persist to $T>t_0$. 
However, the results suggest that the double-peak structure in $C_v$ survives at this disorder strength
with the high-temperature peak signaling moment formation and the low-temperature peak signaling 
moment ordering.

For the energy, the entropy and the structure factor in Figs.~\ref{fig:initial}(a), 
\ref{fig:initial}(d), and \ref{fig:initial}(f) the convergence in $m$ seems even faster. 
The fact that the double occupancy in Fig.~\ref{fig:initial}(b) does not show as fast of a 
convergence to the continuous disorder limit at $T<t_0$ 
compared to the other quantities can be understood based 
on the fact that double occupancy is a local quantity, and in this case, the disorder is also on 
local energies. As we will see later, the converged values of this property are also the most
affected by increasing the strength of the disorder in the chemical potential.

 \begin{figure}[t]
	\includegraphics[width=1\linewidth]{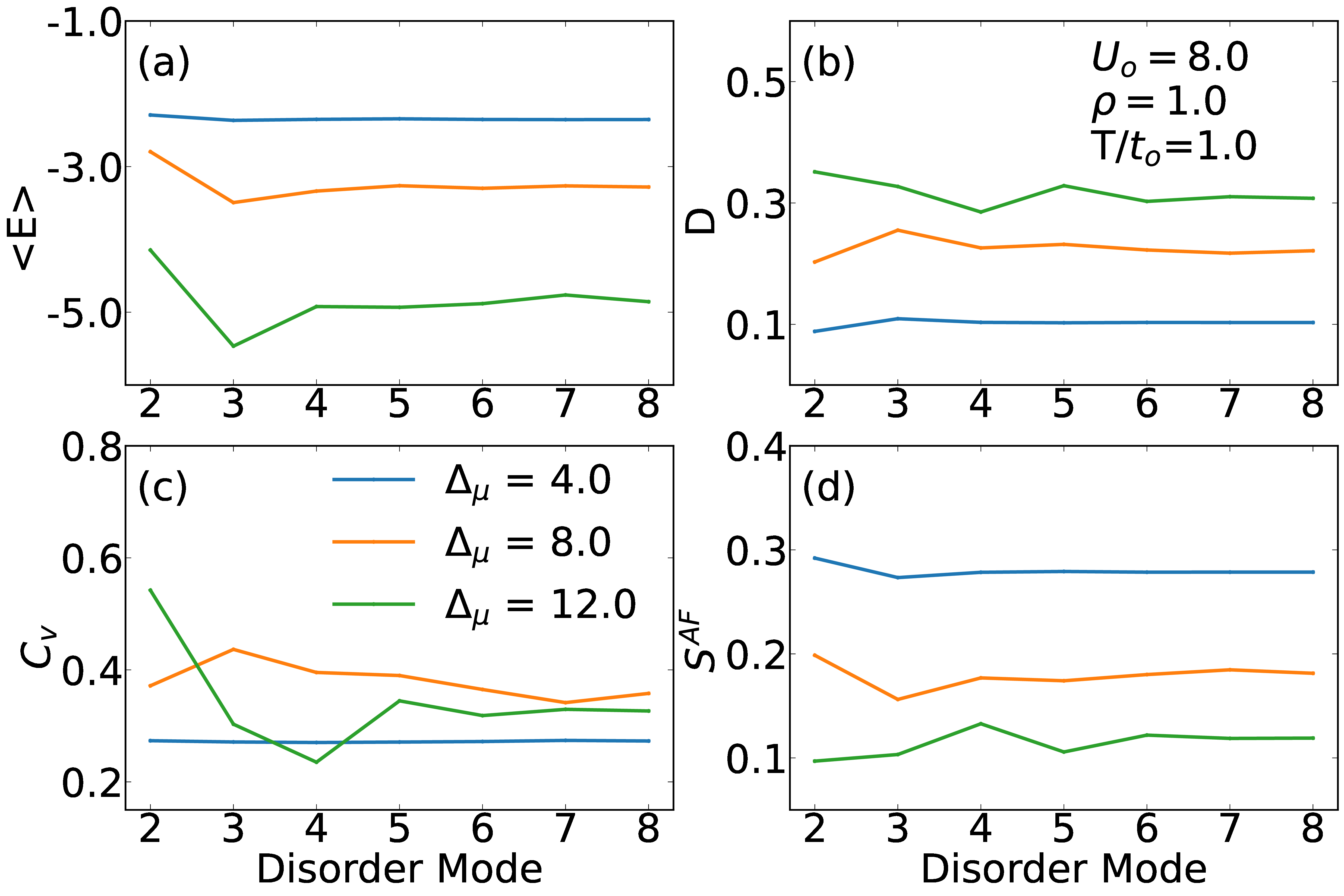}
	\caption{\label{fig:vsmdiffDmu} Convergence in the number of disorder modes for various 
	chemical potential disorder strengths. Same (a) average energy, (b) double occupancy, 
	(c) specific heat, and (d) entropy as in Fig.~\ref{fig:initial} at half filling as a function of 
	the number of disorder modes. All properties are evaluated 
	at $T/t_0=1$, and for $U_0=8$. The values are from NLCE results in the fifth order. }
\end{figure}

Clearly, the convergence in the number of modes depends on the quantity 
under investigation, the temperature, and as one might expect, the strength of the disorder itself. 
To put these in perspective, in Figs.~\ref{fig:vsm} and \ref{fig:vsmdiffDmu}, 
we plot four representative thermodynamic quantities from Fig.~\ref{fig:initial} as a function
of $m$ at various temperatures, or at a fixed temperature for various $\Delta_\mu$. 
Here, we have included both odd and even $m$ up to $m=8$. As one can see in 
Fig.~\ref{fig:vsm}, after relatively large initial variations from bimodal to 
three and four modes, most quantities quickly saturate to final values at $T\gtrsim t_0$
while that is not the case for every quantity at $T<t_0$ 
[see e.g. $C_v$ in Fig.~\ref{fig:vsm}(c)]. In Fig.~\ref{fig:vsmdiffDmu}, we
can see that the fluctuations in properties over different values of $m$ increases 
as the disorder strength increases. Nevertheless, these results suggest
that if the NLCE is converged, the approach towards the continuous disorder limit
is quite fast for the energy and several other properties of the Hubbard model 
and can be achieved within four to six modes. Care must be taken when 
it comes to other properties, such as the heat capacity, when the disorder strength 
is relatively large. We emphasize that
converged NLCE results are valid in the thermodynamic limit and do not contain
any finite-size errors.

\subsection{Disorder in the chemical potential at half filling}

\begin{figure}[t]
	 \centering
	\includegraphics[width=1\linewidth]{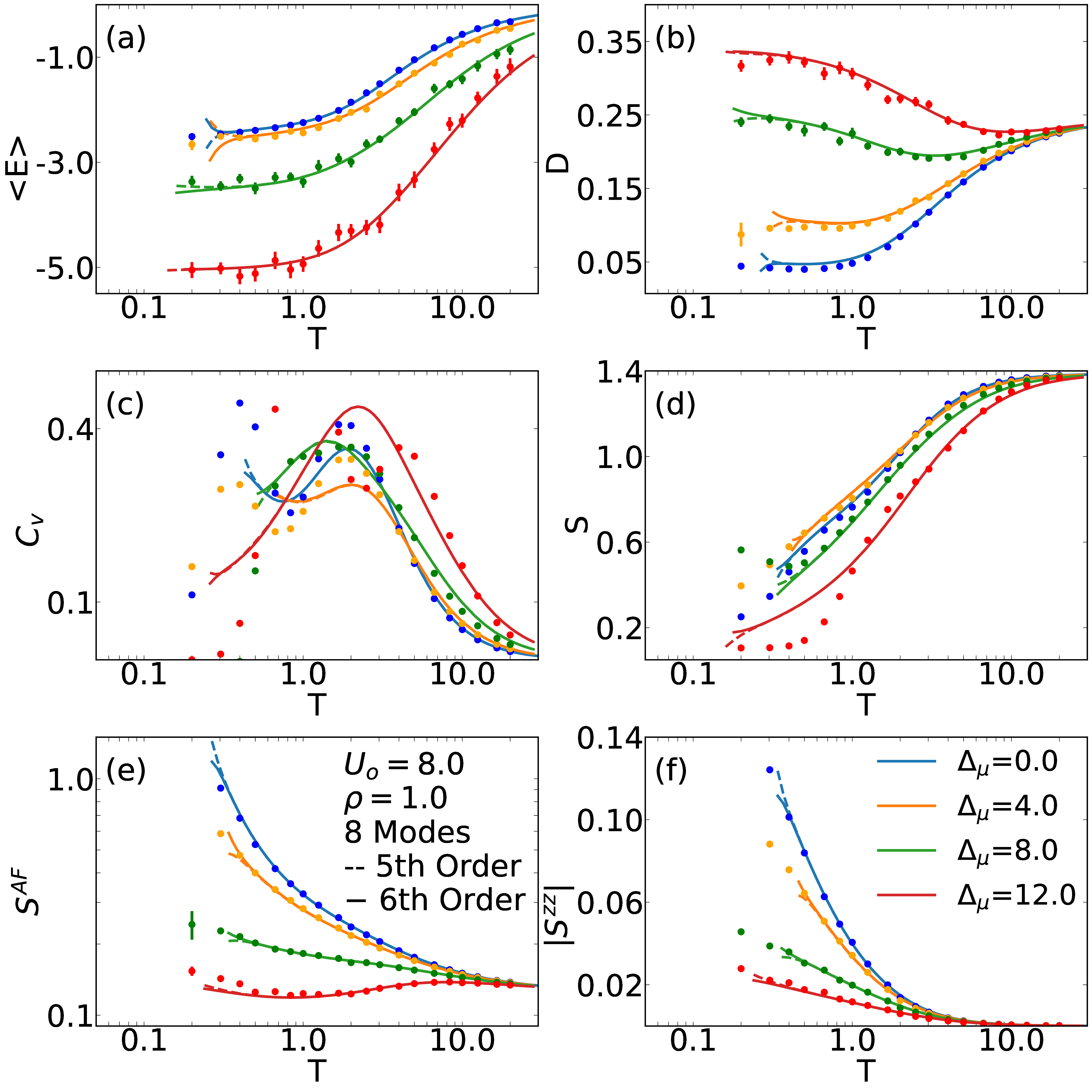}
	\caption{\label{fig:diffDmu} Comparison of NLCE results for various chemical potential disorder 
	strengths vs temperature. Here, the same quantities as in Fig.~\ref{fig:initial} are shown 
	for an eight-mode disordered system at half filling with an interaction strength of $U_0=8$. 
	Shown are the fifth (dashed lines) and sixth orders (solid lines), 
	except for the clean system, where eighth and ninth orders are used.
	Included are DQMC solutions denoted by colored dots. They are for a $10\times 10$-site lattice 
	and averaged over at least 100 disorder realizations. Unless shown, symbol sizes 
	indicate an upper bound for statistical error bars in all figures. The color mapping for the disorder 
	ranges are the same as for the NLCE solutions.}
\end{figure}

Through the evolution of our thermodynamic properties upon changing the strength of the potential 
disorder, we find that the Hubbard system changes character when the disorder strength exceeds that
of the local repulsive interaction. In Fig.~\ref{fig:diffDmu}, we show the same quantities as in 
Fig.~\ref{fig:initial} obtained from six orders of the NLCE for $m=8$ and four different values
of the $\Delta_\mu$ from 0, representing the clean system, to $12t_0$. 
Increasing $\Delta_\mu$ causes the fraction of doubly occupied sites to increase dramatically, 
even beyond the uncorrelated value of $1/4$ as the temperature decreases, while the average energy
is greatly suppressed. The penalty for double occupancy is now in the range $[U_0- \Delta_\mu,U_0+ \Delta_\mu]$.
Therefore, a strong disorder in the chemical potential, comparable to, or larger than 
$U_0$, means having sites with a negative potential, favoring double occupancy to 
lower the overall energy. For weak disorder, the enhancement in double occupancy (suppression of
moment formation) broadens the high-temperature peak in $C_v$. It is also expected to 
broaden the low-temperature peak as the magnetic energy scale, $J=4t_0^2/U_0$ for 
the clean system, will assume a range too~\cite{m_ulmke_98}.

The behavior is similar to what 
has been previously seen in QMC simulations of the disordered Hubbard model~\cite{t_paiva_15,m_enjalran_01,m_ulmke_98},
and is also consistent with the observation that repulsive interactions larger than the disorder strength can 
take the system out of the disorder-induced insulating phase away from half filling~\cite{p_denteneer_99,s_kondov_15}.
In the clean $\Delta_\mu=0$ limit, Qin {\it et al.}~\cite{m_qin_16} find ground-state double occupancy 
values extrapolated to the thermodynamic limit that are in close agreement with our NLCE results 
at the lowest temperatures.

In Ref.~\cite{m_ulmke_98}, it is shown that for $U_0=4$, the antiferromagnetic correlations disappear 
when $\Delta_\mu\sim 10$ on a $4\times 4$ cluster, which is also consistent with the 
trends seen for the magnetic correlations in Figs.~\ref{fig:diffDmu}(e) and \ref{fig:diffDmu}(f), 
as well as in Figs.~\ref{fig:U4}(e) and \ref{fig:U4}(f) below ($U_0=4$),
as the disorder strength increases. We note that our NLCE results are exact and in the 
thermodynamic limit, apart from possibly a few percent error for $C_v$ at the lowest 
temperatures shown due to the finite number of disorder modes.

For comparison, we have reproduced DQMC results for the clean and disordered systems for 
quantities shown in Fig.~\ref{fig:diffDmu}. The circles in Figs.~\ref{fig:diffDmu}(a),
\ref{fig:diffDmu}(b), \ref{fig:diffDmu}(e), and \ref{fig:diffDmu}(f) are obtained 
directly using DQMC for a $10\times 10$ lattice after averaging expectation values 
over at least a hundred random disorder realizations. We find a very good 
agreement between those and our NLCE results in the thermodynamics limit considering that 
there may still be some finite-size and Trotter systematic errors present in the DQMC results. We
find that keeping the number of realizations fixed, the fluctuations
in the DQMC data increases as $\Delta_\mu$ increases. It is worth pointing out that
NLCE is a far more efficient method for this problem, in terms of the computational cost,
than the DQMC on the $10\times 10$ cluster. A single run of the latter, sweeping all the temperatures
shown, takes between 6,800 and 34,000 CPU hours depending on the imaginary time step, 
whereas the same calculations on an arbitrarily fine temperature, 
or even chemical potential grid, takes about 600 CPU hours using
Intel E5-2680v4 processors in our computer cluster.

The circles in Figs.~\ref{fig:diffDmu}(c) and \ref{fig:diffDmu}(d)
are obtained through the following fit of the DQMC energy~\cite{c_huscroft_99}
\begin{equation}
E=E_0+\sum_{j>0}c_j e^{-j\delta/T},
\end{equation}
where $E_0$, $c_j$ and $\delta$ are the fitting parameters.
The specific heat is then readily available and the entropy is obtained after integration as 
\begin{equation}
S=S_0 + E_0/T - \sum_{j>0} c_j \frac{1-e^{-j\delta/T}}{j\delta},
\end{equation}
where $S_0=\ln 4$ is the half filling entropy per site at infinite temperature. We keep five 
terms in the series for the energy and perform a least-square fit using values at 20 
temperatures on a logarithmic grid between $T\sim 0.2t_0$ and $T\sim 20t_0$.
We note that the fit, and hence, the estimate for $C_v$ and $S$ can be
systematically improved by increasing the number of grid points in temperature
and/or reducing the statistical errors by considering more disorder 
realizations. So, the DQMC results shown do not necessarily represent the 
most accurate ones that can be achieved, rather, those from typical 
reasonable calculations at $T\ge 0.2t_0$. In fact, we find it difficult to 
obtain good agreement between NLCE and DQMC for $C_v$ (and also often for $S$) below 
a temperature of the order of $t_0$ due to poor fits to the energy, although the two methods
generally agree on the trends. A counter example is 
the DQMC result in Fig.~\ref{fig:diffDmu}(c) suggesting a double-peak structure in 
$C_v$ for $\Delta_\mu=12t_0$, which is not supported by the NLCE results.

The change of character of the system as $\Delta_\mu$ increases is visible in the transformation 
of the specific heat and the entropy as well. We know that $C_v$ for the clean Hubbard model at half filling
has two distinct peaks corresponding to moment formation at $t_0<T<U_0$ and moment ordering 
at $T<t_0$~\cite{t_paiva_01,J_bonca_03}. The blue curve in Fig.~\ref{fig:diffDmu}(c) for the clean 
system clearly captures the high-temperature peak and hints at the appearance of another 
low-temperatures peak, observed in previous NLCE studies of the clean model using higher
orders in the expansion~\cite{e_khatami_12b}. Upon the introduction of a small disorder in 
$\mu$ with a strength of $4t_0$, the high-temperature peak loses some weight while its location remains
largely unaffected; moment formation remains the dominant physics. 
The less prominent peak is a signal for additional structural changes in $C_v$
appearing in the low-temperature region associated with magnetic ordering in 
the clean system. However, increasing $\Delta_\mu$ further increases the weight for the 
high-temperature peak again and even creates additional weight at temperatures above the peak.
For $\Delta\mu\gtrsim U_0$, the peak is no longer associated with 
moment formation, but instead with particles localizing at sites with the lowest chemical potentials.
Eventually, for a large enough $\Delta_\mu=12t_0$, almost the entire weight seems to fall under
one high-$T$ peak. This is also reflected in the rapid quench of the entropy at this $\Delta_\mu$
and signals that the system is quickly settling into this phase. As can be seen in Fig.~\ref{fig:diffDmu}(e) 
and \ref{fig:diffDmu}(f), any magnetic correlations are also greatly suppressed, 
a behavior previously shown by Ulmke et al.~\cite{m_ulmke_95,m_ulmke_97,m_ulmke_98_b} and 
Enjalran et al.~\cite{m_enjalran_01}.

\subsection{Disorder in the repulsive interaction at half filling}

\begin{figure}[t]
	 \centering
	\includegraphics[width=1\linewidth]{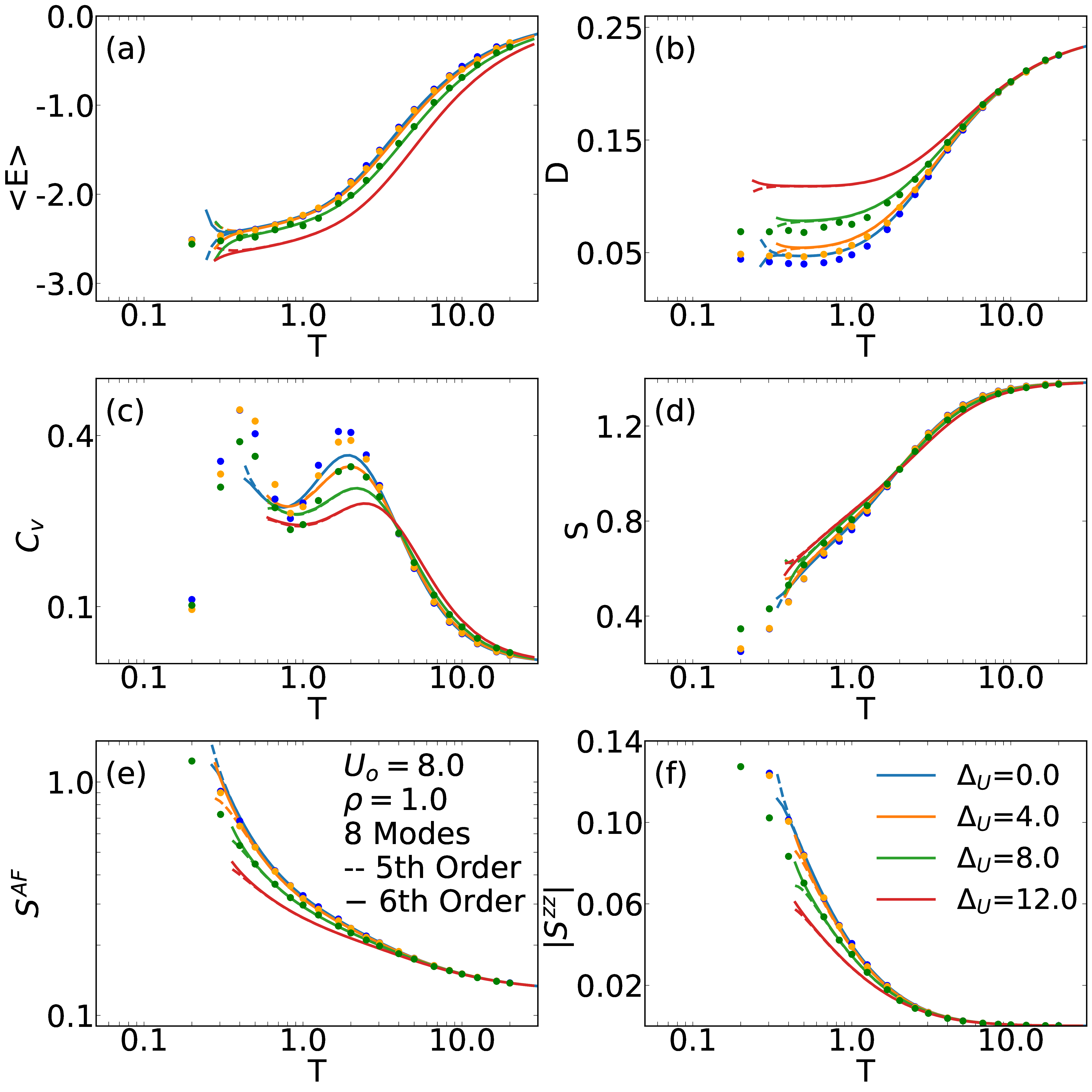}
\caption{\label{fig:diffDU} 
Similar to Fig.~\ref{fig:diffDmu}, but for various disorder strengths in the interaction potential.} 
\end{figure}

We find that the effect of disorder in the interaction strength, which can be seen for 
our thermodynamic properties in Fig.~\ref{fig:diffDU}, is less dramatic than the effect 
of disorder in the chemical potential. 
We have chosen the same four values for $\Delta_U$ as for $\Delta_\mu$ in Fig.~\ref{fig:diffDmu}.
We find that the double occupancy retains its character for $\Delta_U$ as large as $12t_0$; 
decreasing as the temperature decreases. However, its low-temperature values steadily increase as
$\Delta_U$ increases as a result of roughly half of the sites having a smaller penalty for double
occupancy than in the clean system. For $\Delta_U>U_0$, a fraction of sites are even expected to favor 
the formation of doublons. We benchmark our results in the case of $\Delta_U=0$, $4$, and $8$ 
against DQMC and find very good general agreements. We observe significant fluctuations 
in $D$ and $E$ from DQMC, which we attribute to an insufficient number of disorder realizations, and a 
systematic deviation from the exact NLCE results in $D$ at low temperatures for DQMC, beyond 
the statistical error bars, which we attribute to the systematic Trotter error in the latter 
[see Figs.~\ref{fig:diffDU}(a) and ~\ref{fig:diffDU}(b)]. The latter can be mitigated through 
extrapolation in the imaginary time step. We note that DQMC calculations for 
$\Delta_U>8$ run into technical difficulties as the interaction can take a negative sign 
(become attractive) on some sites, and so the corresponding results have not been obtained.

Similarly to the case of disorder in the local potential, the energy also decreases, yet less rapidly, 
upon increasing $\Delta_U$, here as a result of reduced repulsive, or attractive, interactions 
on sites that are most likely to be doubly occupied. 
The heat capacity and the entropy in Figs.~\ref{fig:diffDU}(c) and ~\ref{fig:diffDU}(d) see 
relatively minor changes upon the introduction of the interaction disorder. The peak in 
$C_v$ broadens and moves to slightly higher temperatures with 
increasing $\Delta_U$, reflecting favoring of doublons and reduction in moment formation
in comparison to the clean system. At lower temperatures, the peak associated with moment
ordering is also expected to broaden since disorder in interaction directly results in 
disorder in $J$.
The magnetic correlations [see Figs.~\ref{fig:diffDU}(e) and ~\ref{fig:diffDU}(f)] are
suppressed with $\Delta_U$, especially when $\Delta_U>4$ as a result of having 
less moments available for ordering. 
We do not find $\Delta_U$ affecting the convergence of the NLCE in significant ways for 
any of the properties we have studied.

\subsection{Disorder in the hopping amplitude at half filling} 

 \begin{figure}[t]
	 \centering
	\includegraphics[width=1\linewidth]{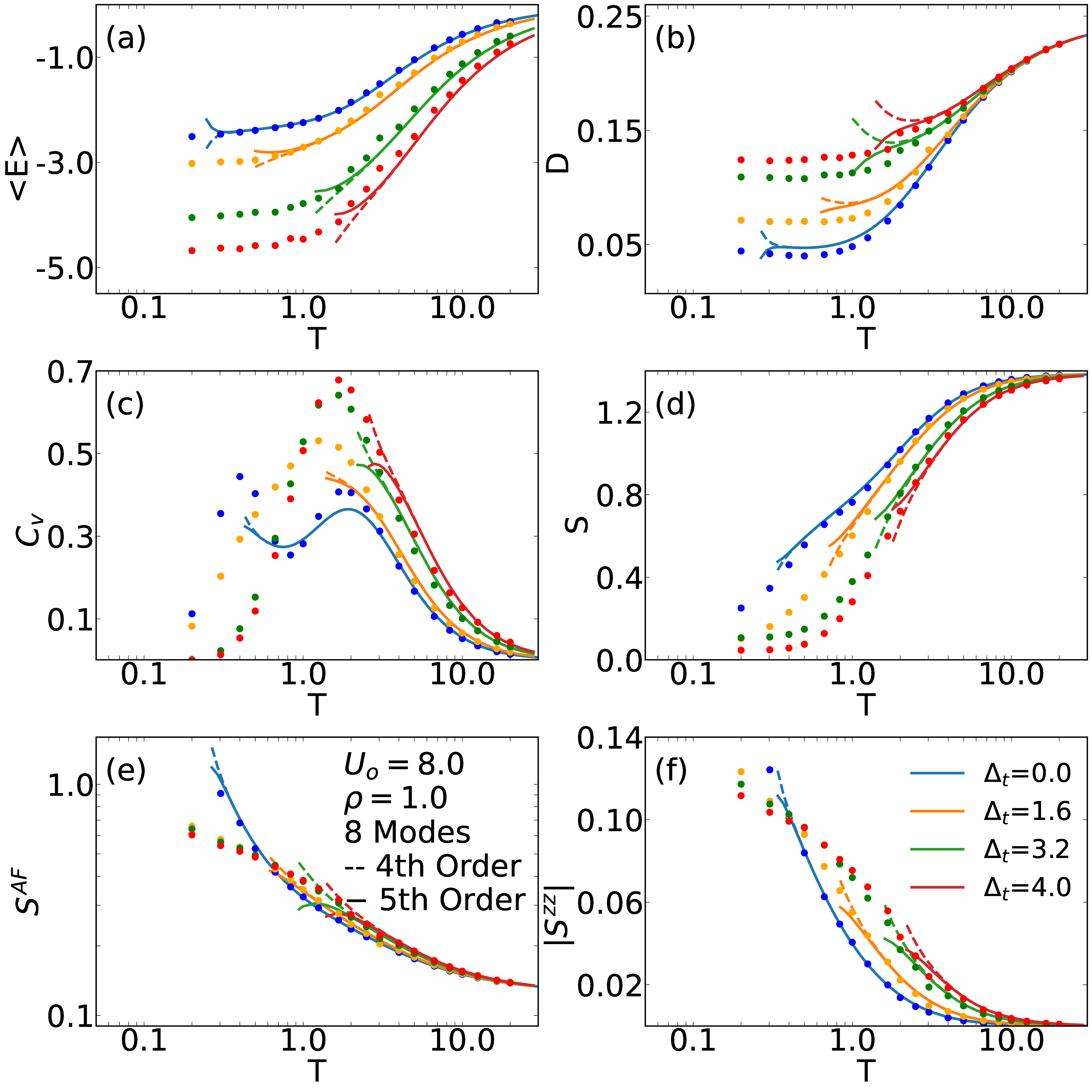}
\caption{\label{diffDt} Similar to Fig.~\ref{fig:diffDmu}, but for various disorder strengths in the hopping integral.}
\end{figure}

We have also considered the model with bond disorder (disorder in the hopping integral $t$). 
As can be seen in Fig.~\ref{diffDt}, in this case, the convergence of the series worsens substantially 
as the disorder strength increases. We attribute this to the non-local nature of this type of disorder 
and the sensitivity of the NLCE's performance to the variation in the non-local correlations of the system.  
However, already at high temperatures results point to interesting physics.
Like in the case of chemical potential disorder, here the bond disorder lowers the 
energy dramatically, but the double occupancy does not increase as dramatically as in the former case 
at intermediate and low temperatures, as can be seen in Figs.~\ref{diffDt}(a) and \ref{diffDt}(b). 
This can be explained through the increased tendency to form singlets on strong bonds 
that experience larger hoppings as the disorder increases, consistent with the observations about the 
magnetic behavior of the system (see below).
Similar to the case of chemical potential disorder, the faster initial drop in the entropy [Fig.~\ref{diffDt}(d)]
and the sign of increase in the high-temperature peak of the heat capacity [Fig.~\ref{diffDt}(c)], 
taking up much of the weight as the disorder increases, also capture the change in physics of the system, in this case
even for disorder strengths as small as $\Delta_t=1.6$.
Interestingly, for the bond disorder strengths we have used, we do not see large 
statistical errors in the DQMC results, which prove very useful in completing the picture 
for the thermodynamics of the system.

The above observations are consistent with the picture drawn by the magnetic correlations shown in Figs.~\ref{diffDt}(e) 
and \ref{diffDt}(f); while short-range correlations get an early boost upon lowering the temperature when the disorder strength increases, 
the antiferromagnetic structure factor, which encompasses  long-range correlations, does not 
experience a big enhancement. The behavior of $S^{zz}$ is unlike that for the system with onsite 
potential or interaction disorder. The latter are detrimental to any type of magnetic correlations, 
whereas hopping disorder can offer weak and strong bonds, favoring singlet formation on the 
strong bonds at the expense of long-range order in the ground state. For this reason, we expect 
that in the presence of sufficiently strong bond disorder $S^{AF}$ will saturate at low temperatures.
In Ref.~\cite{m_ulmke_97}, DQMC results for $U_0=4$ show that the normalized structure factor at $T=0.1t_0$
approaches its uncorrelated value on a $10\times 10$ cluster for $\Delta_t\sim 1.5$.

\subsection{Away from half filling}

 \begin{figure}[t]
	 \centering
	\includegraphics[width=1\linewidth]{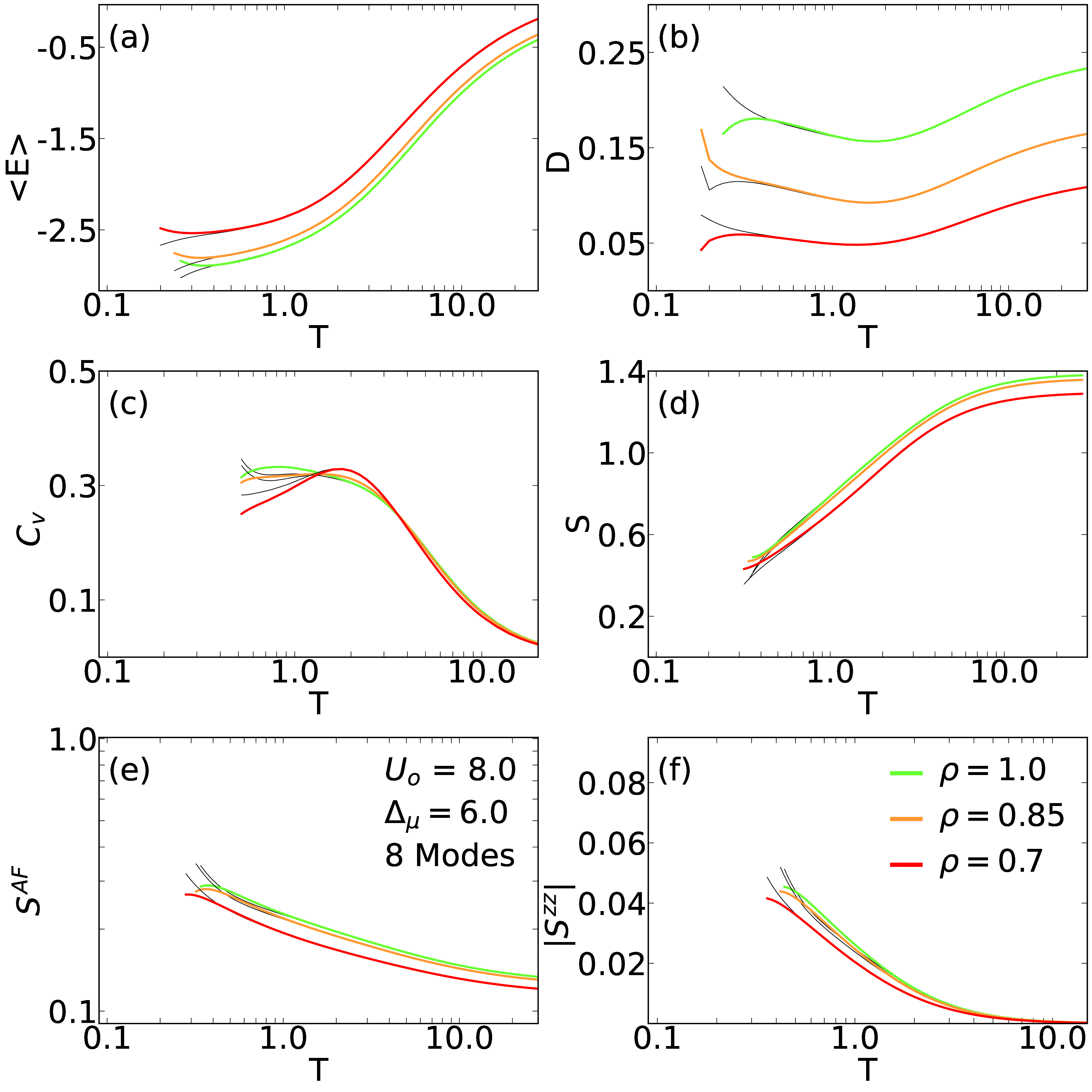}
\caption{\label{diffDmu_away} 
Same properties as in Fig.~\ref{fig:diffDmu}, but at a fixed $\Delta_{\mu}=6$ away from half filling.
Fourth order solutions are presented as black lines and fifth order as colored.}
\end{figure}

Our method can access thermodynamic properties and 
correlations functions of the system in the thermodynamic limit not only at half 
filling, but also all the other densities in a single run. Choosing a fine grid for the 
chemical potential allows for the numerical conversion to fixed densities for 
disordered systems after the average over realizations is performed. Here, we present 
results away from half filling for the two cases of $\Delta_\mu=6$ and $\Delta_U=6$ 
in Figs.~\ref{diffDmu_away} and \ref{diffDU_away}. They are similar to Figs.~\ref{fig:diffDmu} 
and ~\ref{fig:diffDU}, except that we have fixed $\Delta_\mu$ and shown results at 
fixed densities of $\rho=1.00, 0.85$ and 0.70 in them.

\subsubsection{Chemical potential disorder}

The convergence of the NLCE properties for the clean Hubbard model away from half filling
is typically lost at higher temperatures than can be achieved at half 
filling~\cite{E_khatami_11b,b_tang_13,e_khatami_15,e_khatami_16}. However, for the 
disordered systems, we observe that away from half filling, this temperature is comparable to, 
or even lower than, the lowest convergence temperature at half filling. With the onsite energy 
disorder (Fig.~\ref{diffDmu_away}), we do not find significant changes in the behavior of the 
quantities as the disordered system with already suppressed magnetic correlations is doped 
away from half filling. However, there is a notable drop in the fraction of doubly occupied 
sites upon decreasing the density, which can be expected since having fewer particles
directly translates to fewer double occupancies~\cite{E_khatami_11b,b_tang_13}.
This effect is reflected in the energy, and in turn in the $C_v$, which starts to develop a high-temperature 
peak when $\rho=0.7$. The peak is close in height and location to that observed for the clean system 
at the same filling~\cite{d_duffy_97} and shows that effects of moderate site disorder 
at high temperature are likely minimal for the dilute system.
The magnetic correlations are also slightly weakened as a result of fewer number of moments in the system.

We are not showing comparisons to DQMC results away from half filling since the combination
of the fermion ``sign problem"~\cite{e_loh_90,v_iglovikov_15} and the presence of disorder are 
expected to introduce large 
error bars at temperatures lower than $t_0$, especially around $\rho=0.875$, also complicating the 
search for the average chemical potentials as a function of temperature that would yield
the correct fixed densities. That would be beyond the scope of our work given that
we have already established the general agreement of our NLCE results with those of DQMC 
for finite clusters at half filling. It also displays an advantage of using the 
NLCE; exact information about all fillings are readily available after a single run.
In fact, such exact results in the thermodynamic limit for the disordered Fermi-Hubbard 
model away from half filling, while missing from the literature to the 
best of our knowledge, are often 
necessary for characterization of fermionic systems in optical lattices since the 
existence of the trapping potential leads to a range of densities.

\subsubsection{Interaction disorder}

 \begin{figure}[t]
	 \centering
	\includegraphics[width=1\linewidth]{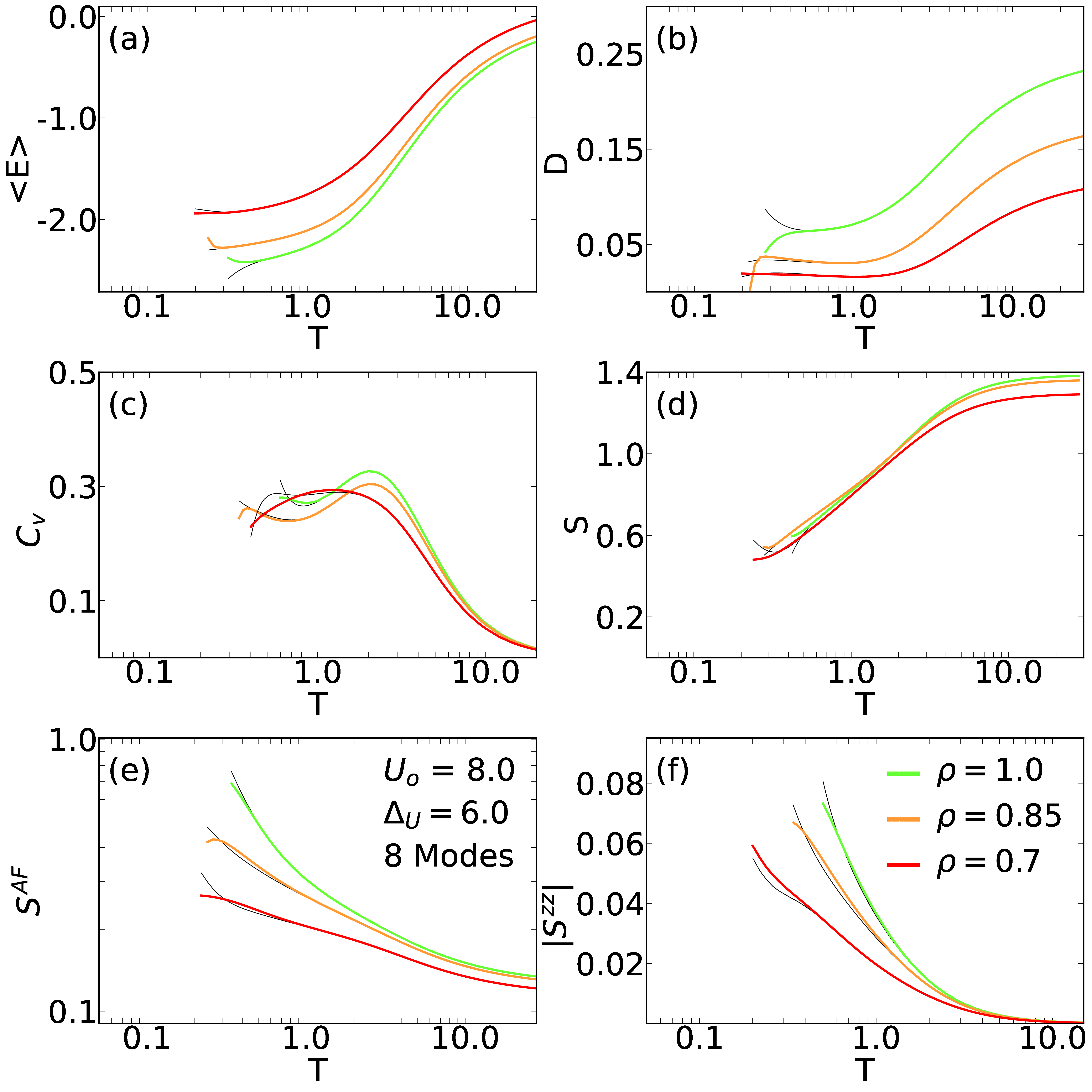}
\caption{\label{diffDU_away} 
Same as in Fig.~\ref{diffDmu_away}, except for an interaction disorder of $\Delta_U=6$.}
\end{figure}

A disorder in interactions with the same strength seems to have a more dramatic effect
on the system and its correlations away from half filling (see Fig.~\ref{diffDU_away}).
An already suppressed double occupancy at half filling is further reduced upon doping, 
and unlike the case of disorder in the onsite energies, here, doping to 30\% seems to 
change the high-temperature character of the specific heat from that 
displaying clear charge peak to one in which the 
suppression of charge and spin degrees of freedom seems intertwined.

We saw in Fig.~\ref{fig:diffDU} that the half-filled 
system is affected considerably less due to the interaction disorder in comparison to disorder
in the chemical potential. However, Fig.~\ref{diffDU_away}(e) and ~\ref{diffDU_away}(f) 
show that the magnetic structure factor and nearest-neighbor correlations quickly give 
up this resistance to change with a finite $\Delta_U$ upon doping, leading to values at $\rho=0.7$ that are close to those 
obtained in Fig.~\ref{diffDmu_away} for a finite $\Delta_\mu$, even at half filling.

\subsection{Chemical potential disorder at other interaction strengths}

\begin{figure}[t]
	 \centering
	\includegraphics[width=1\linewidth]{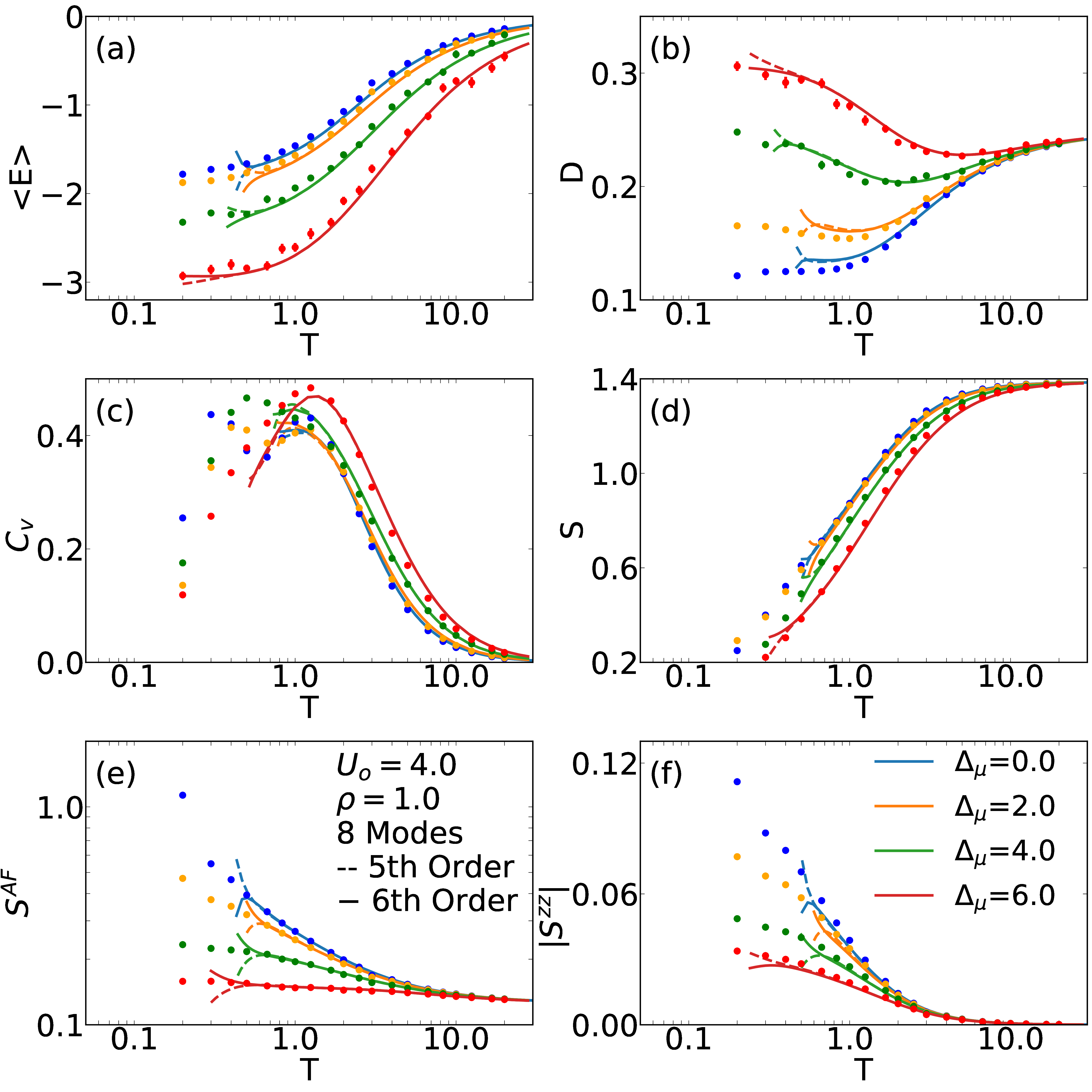}
\caption{\label{fig:U4} Same as Fig.~\ref{fig:diffDmu}, but for $U_0=4.0$. We have 
taken the disorder average in DQMC (circles) over 50 realizations in this case.}
\end{figure}

\begin{figure}[t]
	 \centering
	\includegraphics[width=1\linewidth]{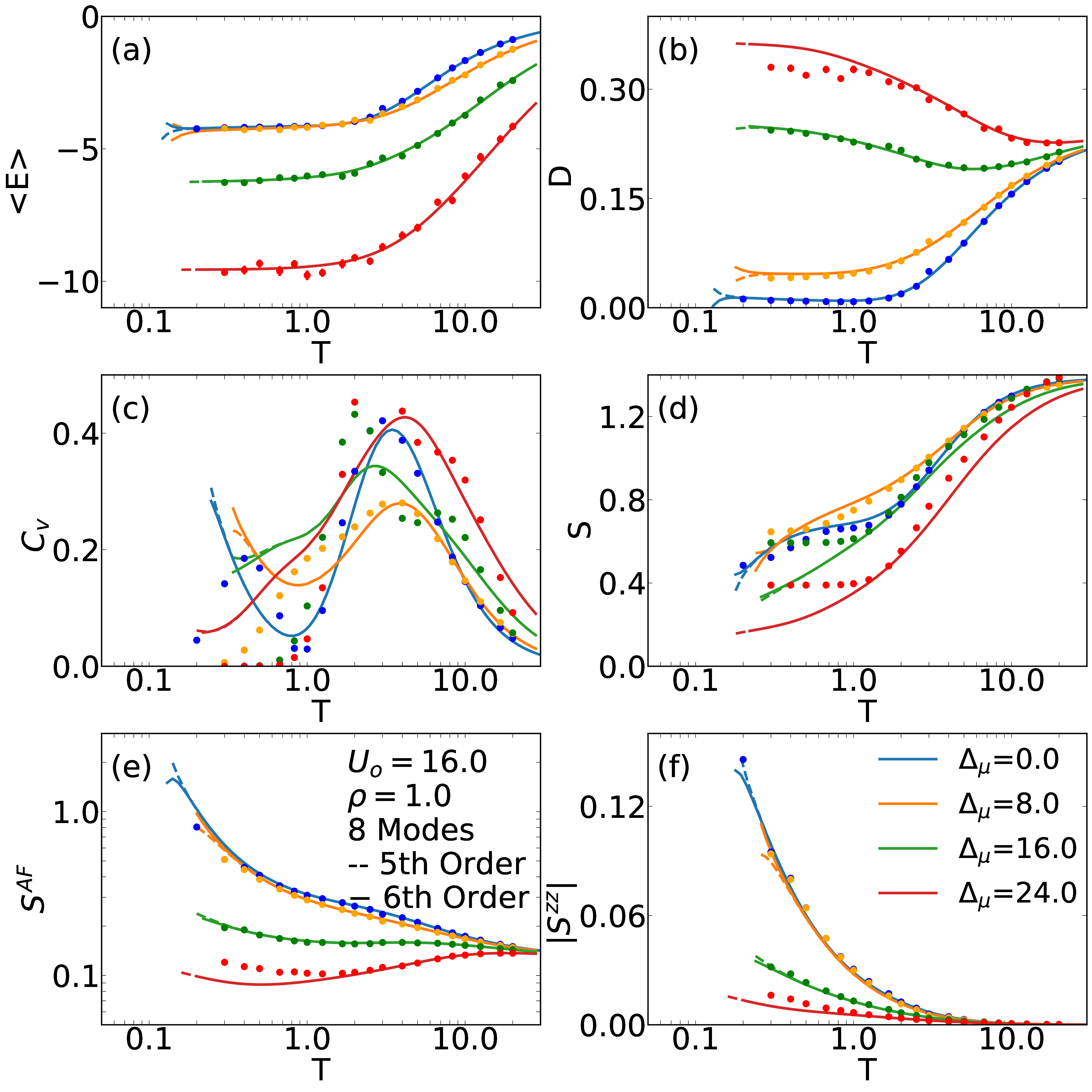}
\caption{\label{fig:U16} Same as Fig.~\ref{fig:U4}, but for $U_0=16.0$.}
\end{figure}

To study the effect of site disorder on the system at other interaction strengths, 
we carry out calculations for the disorder in the 
chemical potential for $U_0=4.0$ and $16.0$ too. These interaction strengths represent
the weak-coupling and the very strongly interacting regions of the clean model, 
respectively. Results are shown in Figs.~\ref{fig:U4} and \ref{fig:U16}.  
When $U_0=4.0$, the Coulomb repulsion and the effective nearest-neighbor
exchange interaction at half filling, $J\sim 4t_0^2/U_0$, are of the same 
order of magnitude. For smaller $U_0$, not only other higher-order terms may have to 
be taken into account, but also the formation of well-defined 
moments in the clean system is largely hindered~\cite{t_paiva_01}. 
This is exacerbated in the presence of weak disorder, as can be inferred from the 
trend in the double occupancy in Fig.~\ref{fig:U4}(b).

We find that many of the trends we observed in the thermodynamic properties  for $U_0=8.0$ 
upon the introduction of disorder in the chemical potential  in 
Fig.~\ref{fig:diffDmu} extend to other values of $U_0$ as well. Most notably,
the fact that the system behaves qualitatively differently as the temperature
is lowered when the strength of the disorder reaches and exceeds the 
interaction strength, e.g., $D$ increasing with $T$, the suppression of 
magnetic correlations, etc. With $U_0=16.0$ in Fig.~\ref{fig:U16}(c), we can 
also clearly observe that the evolution of the specific heat as a function of temperature 
with increasing the disorder strength follows the same trends as in $U_0=8.0$.
That is, the high-temperature peak initially takes less of the overall weight as
$\Delta_\mu$ increases, and as $\Delta_\mu$ passes $U_0$, $C_v$ displays 
one broad high-temperature peak, signaling the onset of particles settling at low energy sites.

We note that, as has been established before for clean systems~\cite{e_khatami_12b}, 
NLCE appears to generally converge to lower temperatures as the interaction strength 
increases for the disordered system. As can be seen in Fig.~\ref{fig:U16}(c), more of the 
nontrivial trends near the low-temperature peak of $C_v$ can be recovered in the NLCE
results when $U_0=16.0$ compared to when $U_0=8.0$ in Fig.~\ref{fig:diffDmu}(c).
In turn, the convergence is lost at higher temperatures for $U_0=4.0$ in Fig.~\ref{fig:U4}(c)
and NLCE does poorly even with the high-temperature peak of $C_v$ when $\Delta_\mu<6.0$.
On the other hand, DQMC, while often yielding reliable results at lower temperatures than
what NLCE can access for $U_0=4.0$, does poorly for $U_0=16.0$, especially for larger values of 
$\Delta_\mu$; other than for the energy [Fig.~\ref{fig:U16}(a)], the trends seen in 
DQMC results for thermodynamic quantities at $T\lesssim 1.0$ when $\Delta_\mu=24$ in Fig.~\ref{fig:U16}
cannot be trusted as they significantly deviate from converged NLCE results. That is despite 
reducing the Trotter error and pushing the calculations to our computational limit in this case.
The same can be said about the specific heat and the entropy for any nonzero $\Delta_\mu$
when $U_0=16.0$ [Figs.~\ref{fig:U16}(c) and ~\ref{fig:U16}(d)]. Hence, the results
in Figs.~\ref{fig:U4} and ~\ref{fig:U16} demonstrate the complementarity of the 
NLCE and DQMC methods, extended to disordered Hubbard systems.

\subsection{Chemical potential and interaction disorders in the 3D Hubbard model}

 \begin{figure}[t]
	 \centering
	\includegraphics[width=1\linewidth]{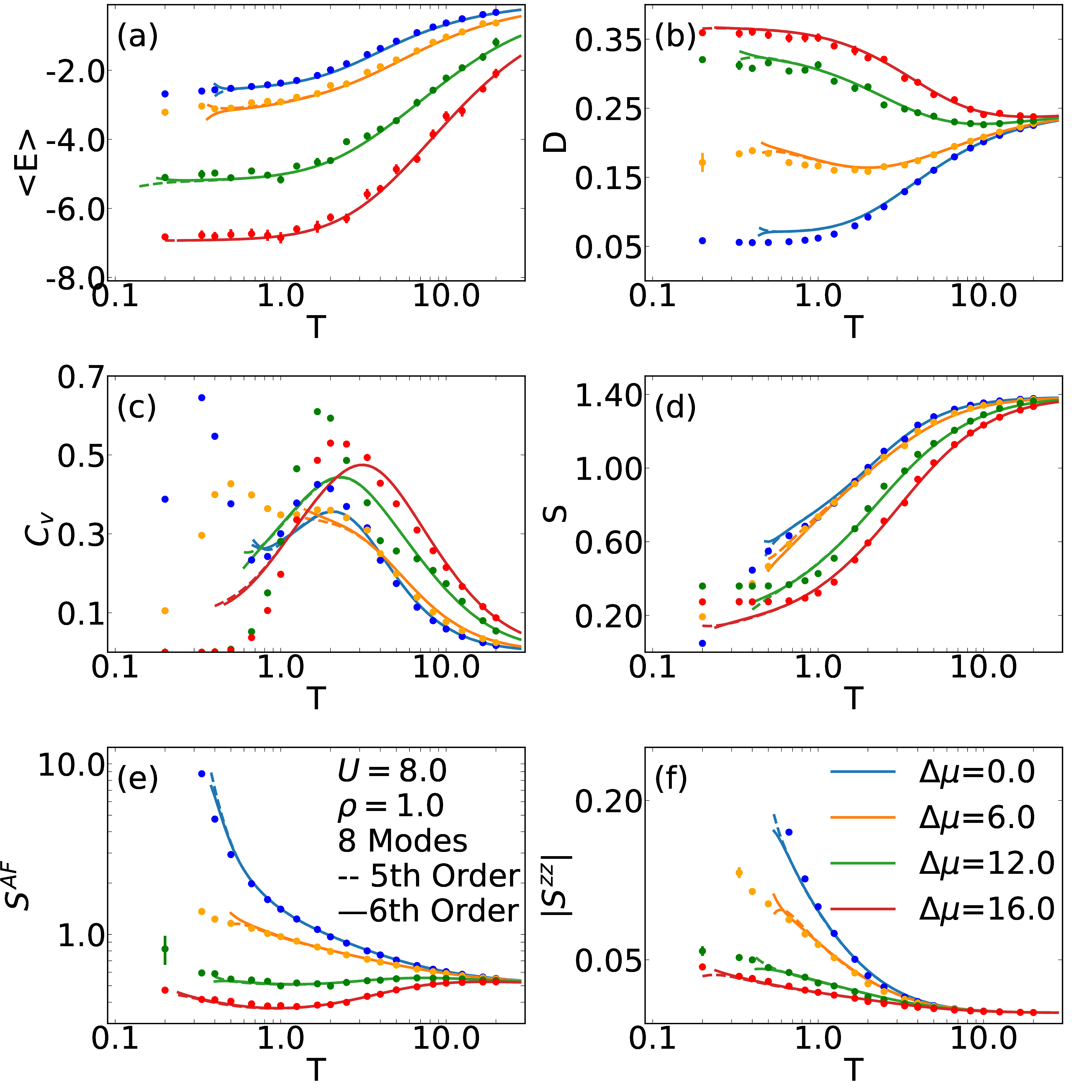}
\caption{\label{diffDmu_3D} Same as Fig.~\ref{fig:diffDmu}, but for the 3D version of 
the Fermi-Hubbard model. Symbols represent DQMC results for a $6\times 6\times 6$ system.}
\end{figure}

By implementing our disorder averaging scheme in the 3D adaptation of the NLCE 
algorithm~\cite{r_hart_15,e_khatami_16}, we have also explored the effect of 
disorder in the chemical potential or the interaction strength on the 
thermodynamic properties of the model in 3D.

In Fig.~\ref{diffDmu_3D}, we show the same properties for the 3D model that are 
shown for the 2D version in Fig.~\ref{fig:diffDmu} when disorder is present in 
the chemical potential for a range of strengths. Here, this range is extended
to $\Delta_\mu=16t_0$ since the 
noninteracting bandwidth is larger in 3D. We find that the trends are similar
to those seen for the system in 2D, except that the magnetic correlations in the 
clean system show a more rapid increase with lowering the temperature and in turn, are
affected more strongly by disorder. The divergent behavior of $S^{AF}$ for the 
clean system in Fig.~\ref{diffDmu_3D}(e) reflects the existence of a finite 
temperature magnetic transition to the Ne\'{e}l phase around $T=0.35t_0$ for 
this interaction strength~\cite{r_staudt_00,e_khatami_16}. Comparisons to 
DQMC results are performed considering a $6\times 6\times 6$ periodic system 
for the latter.

 \begin{figure}[t]
	 \centering
	\includegraphics[width=1\linewidth]{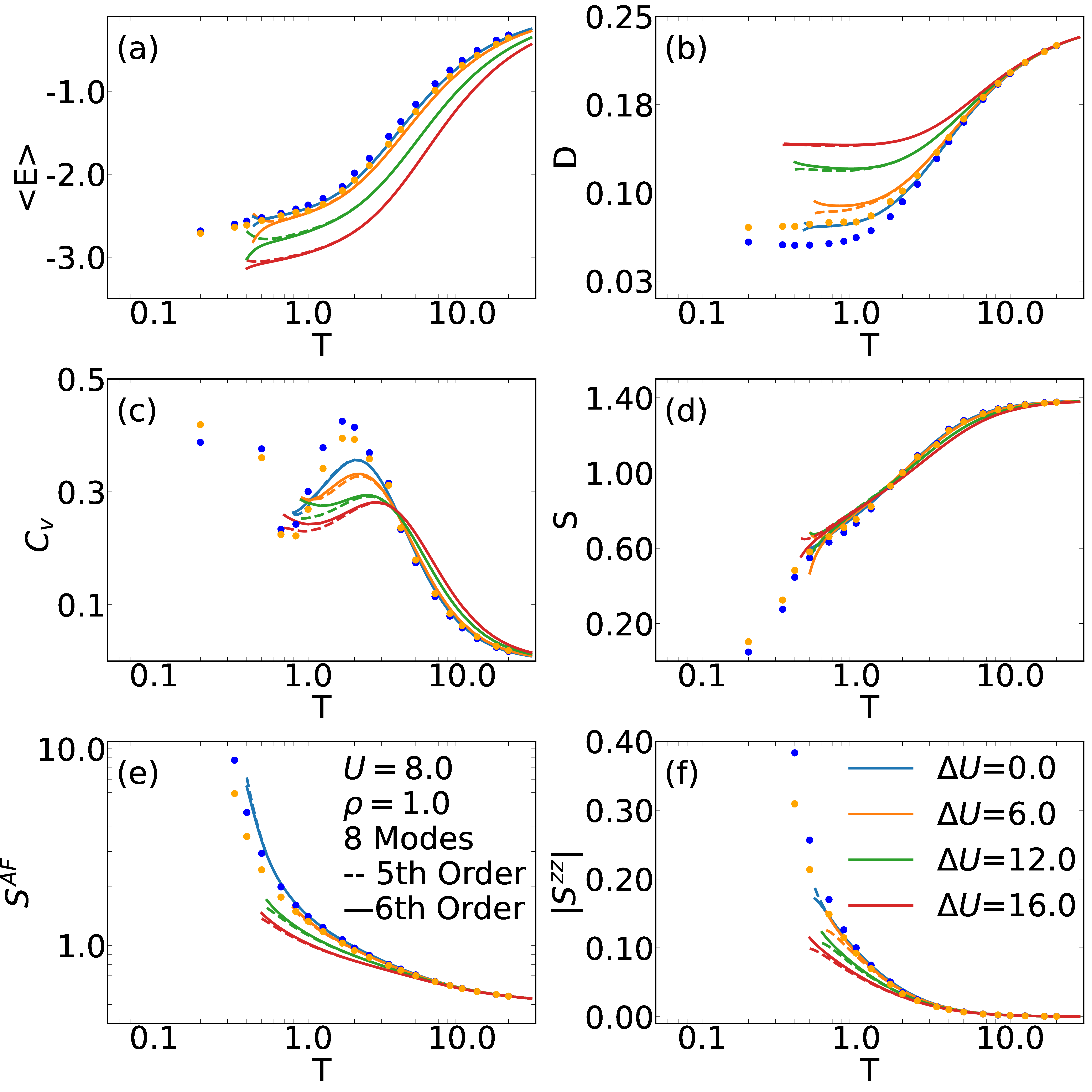}
\caption{\label{diffDU_3D} Same as Fig.~\ref{diffDmu_3D}, except that the disorder 
is in the interaction strength.}
\end{figure}

In Fig.~\ref{diffDU_3D}, we present results for the effect of the interaction disorder on the 3D system 
and find that the same trends we observed for the 2D system largely hold here as well; despite 
a reduction in moments in the disordered system, leading to smaller magnetic correlations, the 
effect of this type of disorder is much less severe, and does not appear to alter the physics of 
the system at temperatures we can access, in comparison to the effect of disorder in the chemical 
potential. Similarly to the 2D case, we cannot carry out the DQMC calculations for $\Delta_U>8t_0$ 
due to technical reasons related to the algorithm. These results show the robustness of 
the NLCE method for studying a wide range of circumstances that include changes in the 
model parameters and dimensionality of the system.

\section{Summary and Discussion}

In summary, we have applied the extension of the NLCE method for studying exact 
finite-temperature properties of disordered quantum lattice models to perform a 
systematic study of thermodynamic properties of the disordered Fermi-Hubbard models 
in two and three dimensions. We consider three different disorder scenarios involving 
the local chemical potential, the onsite interaction strength, and the hopping 
amplitude and monitor the evolution of finite-temperature properties as the strength of the disorder is increased. 
Among other things, we present arguments, based on the trends seen in the heat capacity, 
double occupancy fraction and spin correlations, about the affect of disorder on 
magnetic correlations and localization effects in the system. We demonstrate that 
the results can be reliably extended to incommensurate fillings. We further 
compare our results at half filling with those
obtained using the DQMC algorithm after disorder realization averaging, gauging 
various systematic and statistical errors in the latter as well as the temperature
limitations of the NLCE in different regions.

We find that the half-filled system changes its character from moment formation and 
ordering as the temperature is decreased to particles localizing at sites with the lowest 
energies when the chemical potential disorder strength exceeds that of the local 
repulsive interaction. This trend persists in the weak- and strong-coupling regions 
as well as in 3D. The effect of bond disorder is also dramatic, and is driven by the 
formation of singlets on strong bonds. However, NLCE's performance is significantly 
affected in that case and we cannot access low temperatures. Disorder in the interaction, 
on the other hand, shows relatively small change in the physics of the system even for 
strong disorder strengths, a behavior we observe in both 2D and 3D.

The algorithm adopted here, enabling the treatment of disorder within the NLCE can be
combined with the implementation of real-time correlations functions in the NLCE for 
systems at equilibrium to calculate dynamical properties, as were done in 
Refs.~\cite{m_nichols_19,j_richter_19}, or other implementations for noequilibrium dynamics 
after a quench~\cite{m_rigol_14, m_rigol_14b,b_wouters_14,k_mallayya_17,k_mallayya_18,i_white_17,j_gan_20,j_richter_20}, 
to extend the study of those dynamical properties to 
disordered quantum lattice models and make better connections to optical lattice experiments~\cite{w_morong_16}.
These experiments have so far operated at elevated temperatures 
ranging from $2t_0$ to tens of $t_0$, well within the region of convergence of the NLCE.
Moreover, our results, generally available at an order of magnitude lower temperatures, 
will be useful for future experiments.

In general, in optical lattice experiments aiming to emulate the disordered Fermi-Hubbard model, 
such as those mentioned in the introduction, disorder unavoidably manifests itself 
simultaneously in the on-site potential, the interaction strength, and 
the hopping amplitude, model parameters that already lack homogeneity over the sample
due to the presence of the confining potential. This poses great challenges for parametrizing 
the experiments through comparisons to theoretical results and exposes the need for further 
development of reliable and unbiased numerical methods to meet those challenges.

\begin{acknowledgments}
We thank Richard T. Scalettar for insightful discussions.
This work was supported by the National Science Foundation (NSF) under Grant No. DMR-1918572.
Computations were performed on Spartan high-performance computing facility at 
San Jos\'{e} State University, which is supported by the NSF under Grant No. OAC-1626645.
\end{acknowledgments}

\end{document}